\documentstyle[11pt,appb]{article}

\newcommand{\etal} {{\em et al.}}
\newcommand{\ie}   {{\em i.e.}}
\newcommand{\eg}   {{\em e.g.}}
\newcommand{\cf}   {{\em cf.}}
\newcommand{\D}    {{\cal D}}
\renewcommand{\P}    {{\cal P}}
\renewcommand{\L}    {{\cal L}}
\newcommand{\half}  {\frac{1}{2}}
\newcommand{\longvec}[1]{\overrightarrow{\!#1}}
\newcommand{\VEV}[1]{\left\langle{#1}\right\rangle}

\begin{document}
\title{DISORIENTED CHIRAL CONDENSATE: \\
THEORY AND PHENOMENOLOGY%
\thanks{Work supported by the Department of Energy under contract
 number DE-AC03-76SF00515.}}

\author{J. D. Bjorken
\address{Stanford Linear Accelerator Center\\
 Stanford University, Stanford, California 94309}}
\maketitle
\begin{abstract}
The basic ideas underlying the production dynamics and search techniques
for disoriented chiral condensate are described.
\end{abstract}
%\PACS{PACS numbers come here}
  
\section{Introduction}

These notes are an abbreviated version of lectures given at the 1997
Zakopane School. They contain two topics. The first is a
description in elementary terms of the basic ideas underlying the
speculative hypothesis that pieces of strong-interaction vacuum with a
rotated chiral order parameter, disoriented chiral condensate or DCC,
might be produced in high energy elementary particle collisions. The
second topic is a discussion of the phenomenological techniques which may
be applied to data in order to experimentally search for the existence of
DCC. 

Two other topics were discussed in the lectures but will not be mentioned
in these notes other than in this introduction. One was a review of the
experimental situation regarding DCC searches. There are so far only two
such attempts.  One has been carried out at the Fermilab TeVatron
collider by test/experiment T864 (MiniMax). Preliminary results, all
negative, have been presented at a variety of conferences\cite{refa}. No new
information is available now, and the interested reader is invited to
consult the references\cite{refb}. The other experiment, WA98, is in the
fixed-target heavy-ion beam at CERN. Again there is no evidence of DCC
production\cite{refc}. The analysis methods are at present being refined by that
group and are different than for MiniMax, because they are blessed with a
relatively large phase-space acceptance. A recent status report is given
by Nayak\cite{refd}.

The other omitted topic is quite separate, and has to do with the
initiative called FELIX. It is a proposal for a full-acceptance detector
and experimental program for the LHC dedicated to the study of QCD in all
its aspects---hard and soft, perturbative and non-perturbative. Much has
happened since Zakopane with respect to FELIX. Especially noteworthy is
the production of a lengthy and detailed Letter of Intent\cite{refe}, which
provides much more detail than was possible in the lectures on what it is
about, and in any case provides guidelines for all LHC experiments on
interesting issues and opportunities in QCD worthy of study. 
Unfortunately, at this time of writing, the initiative has run into
difficulties with the CERN committees and management, with its future at
present uncertain.

\section{Theory of DCC Production}
\subsection{Light-quark effective theory (LQET)} 

We accept without question that the usual QCD Lagrangian provides a
correct description of the strong interactions. Nevertheless, at large
distances the spectrum of the theory is that of colorless hadrons, not
colorful quarks and gluons. The confinement mechanism responsible for this
situation is only part of the story. In addition there is an approximate
chiral $SU(2)_L \times SU(2)_R$ flavor symmetry which is spontaneously broken. 
The pions are the collective modes, or Goldstone bosons, associated with
this phenomenon of spontaneous symmetry breaking. In addition, in this
low-energy world where hadron resonances are a dominant feature, the
constituent quark model works quite well, with an interaction potential
which does not involve in any explicit way the gluons; direct gluonic
effects seem to be quite muted\cite{reff}. 

There are a variety of low-energy effective Lagrangians which are in use,
associated with this regime. And they are quite well-motivated, with a
starting point being the fundamental QCD short-distance Lagrangian. The
procedure of starting at short distances and ending up with a
large-distance effective theory depends strongly upon taking into
consideration the effects of instantons\cite{refg}. These lectures are not the place
to go into what instantons are, and it has to be assumed that the
uneducated reader will search elsewhere to find out\cite{refh}. It is rather clear on
phenomenological grounds, and is supported by detailed variational
calculations, that the important instantons have a size characterized by a
momentum scale of about 600 MeV and that the size distribution is quite
sharply peaked about this value\cite{refi}. The instantons form a dilute liquid in
(Euclidean) space-time, with a packing fraction of only 10\%. 
Nevertheless, the interactions between them, and the effects of the
instantons on the Fermi sea of light quarks, are very important. There are
light-Fermion ``zero modes" surrounding the instantons, and their
correlations are such as to rearrange the Fermi sea of light quarks in
just the right way to create the chiral symmetry breaking.

Assuming that these instantons are indeed the most important gluonic
configurations at the 600 MeV scale, their main effect when ``integrated
out" of the QCD path integrals, is to leave behind an effective action of
the Nambu-Jona-Lasinio type between the light quarks.  This effective
action, to be applied at momentum scales below 600 MeV, does indeed imply
spontaneous chiral symmetry breaking and the existence of the pionic
Goldstone bosons, which emerge as composites of the quark-antiquark
degrees of freedom. It also constitutes a definite starting point in
general for constituent-quark spectroscopy.  An extensive amount of work
utilizing this effective action is quite successful phenomenologically\cite{refi}. 

At still lower momentum scales, or larger distance scales, the
constituent-quarks themselves can be ``integrated out" of the effective
action. They are replaced by the pionic degrees of freedom, comprising the
lowest mass scale, or largest distance scale, in the strong interactions.
The effective action looks then very much like the one used for the Higgs
sector. However the action of this effective theory need not be restricted
to be renormalizable.  There will be, in addition to the quadratic
free-particle term and quartic interaction, terms of higher polynomial
order, some with derivative couplings depending upon the choice of
description. This is just the purely chiral effective action studied in
great detail by Gasser and Leutwyler, among others\cite{refk}.

Simplified versions of the chiral effective theory are the linear and
nonlinear sigma models. The linear sigma model is what is isomorphic to
the usual Higgs theory; the nonlinear sigma model essentially "integrates
out" the massive sigma degree of freedom (analogous to the massive Higgs
boson of standard-model electroweak theory), leaving only the pionic
degrees of freedom in the effective action. In what follows, these
effective actions are what are relevant for the description of DCC
because, as we shall see, the space-time scale for the DCC system is
assumed to be ``macroscopic", \ie\  have dimensions large compared to 1
fermi. However, we should emphasize here that the use of sigma models, be
they linear or nonlinear, is only a rough approximation to the true
situation at best, because the Yukawa couplings of the Goldstone system of
pions and sigma to constituent quarks are large and should not be ignored,
except perhaps at the very lowest momentum scales.
\newpage

\subsection{DCC Production Scenarios}

As we have mentioned in the introduction, DCC is by assumption a
``macroscopic" region of space-time within which the chiral order parameter
is not oriented in the same direction in the internal $O(4) = SU(2) \times SU(2)$
space as the ordinary vacuum. If we use for simplicity the language of the
linear sigma model, the sigma and pion fields form a four-vector \boldmath$\phi$ 
\unboldmath in
the $O(4)$ space, and in the vacuum there is a nonvanishing value of \boldmath$\phi $
\unboldmath  in
the $\sigma$ direction.

What we shall assume is that in a high energy collision, there are regions
of space-time where $\langle\mbox{\boldmath$\phi$}\unboldmath \rangle$, essentially a classical 
quantity, is rotated
away from the $\sigma$ direction. At late times, of course, this ``chirally
rotated vacuum" must relax back to ordinary vacuum. The mechanism will
clearly be the radiation of the collective modes, the pions, via a
semiclassical mechanism. It should be clear that, if indeed this happens,
it is quite an interesting phenomenon, because it would provide a rather
direct look into the properties of the QCD vacuum itself.

How might the DCC be produced? In hadron-hadron collisions, most models of
particle production, be they stringy or partonic, put the bulk of the
space-time activity near the light cone. That is, the flow of produced
quanta is concentrated in a rather thin shell expanding from the collision
point at the speed of light. But what happens within the interior of this
shell? If it rapidly relaxes to vacuum, then it is hard to see why the
normal vacuum must be chosen, because the interior region is separated
from the exterior true vacuum by a ``hot" shell of expanding partons.  And
as we shall see, the time scale for the true vacuum to be selected via the
small chiral-symmetry-breaking effects associated with nonvanishing pion
mass is quite long.

It is this ``Baked-Alaska" scenario which will be the main thrust of this
discussion\cite{refl}. However, there is a second space-time scenario relevant to
heavy ion collisions. The idealized case is that of infinite pancakes
colliding with each other at the speed of light, with quark-gluon plasma
produced in between the pancakes at early times. But at late times, after
this plasma goes through the deconfining/chiral phase transition, the
possibility of DCC formation also exists\cite{refm} and has been rather
extensively studied\cite{refn}. The geometry in this idealized situation is that of a
boost-invariant 1+1 dimensional expanding system, and is the most
tractable case to consider from a calculational point of view. We shall
review a simple example studied by Blaizot and Krzywicki in the next
section. 

Can one actually predict that DCC should be produced under these
circumstances? Certainly not. On the other hand can the possibility of DCC
production be excluded theoretically? Again, certainly not.  The subject
needs to be data driven. This is the reason that both Cyrus Taylor and I
decided to go into experimental physics and make an experimental search at
the TeVatron collider. While the MiniMax experiment which emerged is very modest,
and while the results so far are negative, the experience gained has been
invaluable in learning how to construct good search strategies, and in
determining what is necessary in order to do a better experimental job in
the future. 

\subsection{Sigma models}

The space-time scenario for DCC production begins at an early proper time,
\ie\ near the light cone, when the energy density becomes low enough that
the chiral order parameter is nonvanishing. The proper time scale here is
plausibly somewhere between 0.3 and 0.8 fermi. At this point one must
assume initial conditions for the chiral field, which then evolves, at the
simplest level of approximation, according to the classical equations of
motion. A natural hypothesis is, in the context of the linear sigma model,
that the chiral field initially vanishes (at least on average).  This
means it is initially on top of the ``Mexican hat", and rolls off the hat
into the region of the minimum. During this initial roll, the linear sigma
model, at the least, should be used for the proper-time evolution of the
vacuum condensate.  As we shall see, this evolution can go on for anywhere
from one to five fermi of proper time. The formalism is as follows. The
chiral fields can be written as
\begin{equation}
\mbox{\boldmath$\phi$\unboldmath} = (\sigma,\longvec\pi) \qquad i = 1,\ldots 4
\label {eq:A}
\end{equation}
or alternatively as a $ 2 \times 2$ matrix of fields
\begin{equation}
\Phi = \sigma +i\longvec \tau\cdot\longvec\pi
\label{eq:B}
\end{equation}
with the Lagrangian for the former case being
\begin{equation}
\L = \half \left|\partial_\mu
\mbox{\boldmath$\phi$\unboldmath}\right|^2 - \frac{\lambda}{4}\, \left(
\mbox{\boldmath$\phi$\unboldmath}\,^2-f^2_\pi\right)^2 \ .
\label {eq:C}
\end{equation}
At late times, when the chiral field is near the minimum of the
Mexican-hat potential radially, but is still rolling around azimuthally,
the nonlinear sigma model can be used. One writes, in the $2\times 2$ notation
\begin{equation}
\Phi = R\, e^{i\longvec\tau\cdot\longvec\pi}
\label{eq:D}
\end{equation}
and freezes out the radial degree of freedom
\begin{equation}
R \approx f_\pi  \ .
\label{eq:E}
\end{equation}
If one further writes as a special case
\begin{equation}
\Phi = f_\pi e^{i\tau_3\pi_3(x)}
\label{eq:F}
\end{equation}
it follows that the pion field which appears in the exponential can be
shown to obey free-field equations of motion. 
\begin{equation}
\Box\, \pi_3(x) = 0 \ .
\label{eq:G}
\end{equation}
This can be extended using a general, global $SU(2) \times SU(2)$ rotation to a
class of solutions (``Anselm class"\cite{refo}): 
\begin{equation}
\Phi \rightarrow U_L\Phi\, U^\dagger_R \ .
\label{eq:H}
\end{equation}
It is important to keep in mind in what follows that the linear sigma
model leads to nonlinear equations of motion, while the nonlinear sigma
model here leads to linear equations of motion. 

While this discussion has been only at the classical level, there has been
a lot of work, especially by those interested in the heavy-ion DCC, in
going beyond this approximation. The state of the art is to incorporate
quantum corrections to the linear sigma model in the mean-field (or
Hartree, or large-N, or ``random-phase") approximation\cite{refp}. This amounts
essentially to inclusion of tadpole contributions to self-energy parts or
geometrical-series bubble summations to appropriate propagators. While
this level of computation is simple in translationally-invariant systems,
it is not at all simple in cases like this, where the space-time geometry
of sources is quite nontrivial. So far, the inclusion of quantum effects
is important quantitatively but has not seriously changed the qualitative
properties. In this discussion we shall remain within the classical
approximation. 

\subsection{The Blaizot-Krzywicki model}

A very easy application can be made of the above considerations\cite{refr}. Consider
boost-invariant solutions of the nonlinear sigma-model equations within
infinitely large heavy-ion disks receding from each other at the speed of
light after a central collision. The equation of motion
\begin{equation}
\left(\frac{\partial^2}{\partial t^2} - \frac{\partial^2}{\partial z^2}\right)\pi = 0 
\label{eq:I}
\end{equation}
leads to the solution
\begin{equation}
\pi = \alpha\, \ell n\, \tau + {\rm const.}
\label{eq:J}
\end{equation}
with
\begin{equation}
\tau^2 = t^2-z^2 
\label{eq:K}
\end{equation}
and
\begin{equation}
\Phi = f_\pi \left(U_L\, e^{i\alpha\tau_3\ell n\, \tau}\, U^\dagger_R\right) \ .
\label{eq:L}
\end{equation}
We see that in general the chiral angle in isospin space precesses with
increase of proper time, provided the left-handed global rotation differs
from the right-handed one. This curious feature depends on the choice of
geometry, however, and does not generalize to the Baked-Alaska scenario. 

These calculations have been extended to the case of the linear sigma
model, with qualitatively similar results.

\subsection{Baked Alaska}

The simplest case of the Baked-Alaska scenario again utilizes the
nonlinear sigma model. Inside the future
light cone one assumes spherical symmetry. In addition, it is assumed that
for $0 < t < T$, with $T$ some ``decoupling time", there is chirally rotated
vacuum within the light cone, \ie 
\begin{equation}
\mbox{\boldmath$\phi$\unboldmath} =  (0, f_\pi,0,0) \ .
\label{eq:M}
\end{equation}
We have taken, without much loss of generality, the chiral orientation to
be purely in the $\pi^0$ direction, because other cases can be obtained
via global chiral rotations. With this hypothesis, the free equation of
motion for the pion field can be used for times $t > T$, using the boundary
conditions for the field at $t = T$. 
When the pion mass is neglected, the form of the solution must be
that of a right-moving plus left-moving pulse:
\begin{equation}
r\pi(r,t) = f(r-t) + g(r+t) \Rightarrow f(r-t)-f(r+t)
\label{eq:N}
\end{equation}
and the shape of the function $f$ is easily found to be a triangular pulse
of width $2T$ and maximum height $f_\pi T$.
\begin{equation}
r\pi(r,t) = \frac{f_\pi}{2}\, \left[(2T-t+r)\theta(2T-t+r)\theta(t-r)
-(2T-t-r)\theta(2T-t-r)\right] \ .
\label{eq:O}
\end{equation}
Thus for times $t > 2T$ there is only a triangular pulse of pion field
(actually $r \pi$) radiated outward at the speed of light, which
comprises the decay products of the initially formed DCC. 

The calculations and formalism for this example have been laid down in
considerable detail in a recent paper\cite{refs}, to which the interested reader is
referred for details. In particular, the calculations have been redone for
nonvanishing pion mass, as well as for the linear sigma model.
An important feature of this class of solutions is that, when one
generalizes them away from the production of $\pi^0$ field by applying a
global $SU(2) \times SU(2)$ rotation, the right-handed and left-handed $SU(2)$
rotations (\cf\ Eq. (8)) must be the same in order that the vacuum at large
times inside the light cone is the same as the ordinary vacuum.

Another way of looking at all this is in terms of sources. The free wave
equation does not hold on the light cone for times $t < T$. It is in that
region that the hot shell of expanding partons or other quanta separates
the DCC vacuum on the inside from the true vacuum on the outside. So one has,
everywhere in spacetime
\begin{equation}
\Box \pi(x) = J(x) \ ,
\label{eq:P}
\end{equation}
with the source $J$ having support only on the light cone. A careful look
at the discontinuity of the vacuum fields across the light cone leads to
the expression
\begin{equation}
J(x) = \frac{\pi f_\pi}{r}\, \delta(t-r)\theta(T-t) \ .
\label{eq:Q}
\end{equation}
Now there is a relation, well-known from the Bloch-Nordsieck treatment of
semiclassical electromagnetic radiation, between the source which creates
a classical field and the spectrum of quanta produced by the source. 
In this case\cite{reft} it is
\begin{equation}
2E\, \frac{dN}{d^3p} = \frac{1}{(2\pi)^3}\ \left|\tilde J(p)\right|^ 2
\label{eq:R}
\end{equation}
where $\tilde J$ is the Fourier transform of  $J$, put on mass shell. 
\begin{equation}
\tilde J(p) = \int d^4x\, e^{ip\cdot x} J(x) \qquad p^2 = m^2_\pi \ .
\label{eq:S}
\end{equation}
It follows that, with pion mass neglected, the expression for $\tilde J$ is 
\begin{equation}
\tilde J(x) = \frac{4\pi^2f_\pi}{|p|} \int ^T_0 dt\, e^{ipt}\sin pt \ .
\label{eq:T}
\end{equation}
One readily checks that this agrees with the spectrum calculated
explicitly from the equations of motion. Note that the spectrum has a
high-momentum tail coming from the delta-function support on the light
cone, but that the typical momentum scale is of order $1/T$.

It is an important feature of this radiation that it is coherent, which
means that for a specific choice of source the multiplicity distribution 
of produced particles is
Poisson-distributed, and that there is no Bose-Einstein enhancement.
However, upon averaging over sources, in particular their chiral
orientations, these simple features undergo essential complications.  

\subsection{Radially Boost-invariant Baked Alaska}

Another simple and interesting example of DCC production has been
considered by Lampert, Dawson, and Cooper\cite{refu} (hereafter LDC). They assume a
Baked-Alaska scenario, within the linear sigma model, which has full
boost invariance. In other words, the chiral fields existing within the
light cone depend only upon the proper time. This scenario cannot be
regarded as realistic, because the particle distribution must look the
same in all reference frames, and hence cannot have a limited spectrum of
energies. They argue that one can extract the physics by sampling the
distribution at large proper times within a small space-time region.
However, this is not a faithful description of what a real piece of
experimental apparatus would see. 

A better strategy in my opinion is to use their solutions up to time T,
and then assume, as was done in the previous section, that there is at
later times no explicit source on the light cone. In other words the LDC
solution at time T is evolved via the source-free linear sigma-model
equations of motion, in order to generate the asymptotic fields. Some
work has been done along these lines\cite{refv}, but not enough to report here.

However, let us return to the simplified version of LDC. At the classical
level the field equations become ordinary coupled differential equations,
and can be easily solved numerically:
\begin{equation}
\left[ \frac{1}{\tau^3}\  \frac{d}{d\tau}\, \tau^3\, \frac{d}{d\tau}+ \lambda
(\sigma^2+\pi^2-f^2_\pi)\right]\, {\sigma\choose\pi} = {f_\pi m^2_\pi\choose 0}
\ .
\label{eq:U }
\end{equation}
However LDC do better, and include the mean-field quantum corrections. It
has been found, however, that the quantum effects do not qualitatively or
even quantitatively make a big difference\cite{refs}. 

What is noteworthy about the solutions is that it takes a rather long
proper time for the initial chiral fields to ``roll" into the minimum of
the potential; the time taken for the
sigma field to settle down to near its vacuum value is about 5 fermi. 
This long proper-time interval, even in the presence of pion mass, would
indicate the credibility of scenarios which create DCC from deep within
the light cone. However, the finite extent of the source on the light
cone needs to be investigated, as well as the effect of considering a
statistical ensemble of initial conditions for the initial ``roll", before
drawing serious conclusions. 

\subsection{Doing better}

All the theoretical attempts we have described are extremely idealized. 
There does exist some numerical simulation work which attempts less
idealized scenarios\cite{refw}, but I think it fair to say that none of it is yet
very near to what is needed for, say, Monte-Carlo input appropriate to
real experimental searches. There are many issues to be addressed.

Perhaps the most important deficiency is that the idealized cases very
likely have far too much symmetry. Looking at the Baked-Alaska scenario
from the source point of view, it is probably the case that in a given
event not only the source strength but also  the chiral orientation of the source
depends upon where in the lego phase-space one is. 
Since the chiral dynamics is spin-zero, the correlations in the lego
phase-space are most probably short-range, of order one to two units. 
This means that on the sphere, near 90 degrees relative to the beam, only
a steradian at a time may have the rather symmetrical structure of the classical
solutions we have considered.

There is another way of looking at this.  Suppose
DCC is produced and observed. It almost by definition will consist of a
cluster of pions of almost identical momenta. This cluster will have a
rest frame, and in that rest frame the classical radiation field
associated with this cluster will in fact have approximate spherical
symmetry. However in general, this frame of reference is related to the
observer's frame of reference by not only a longitudinal Lorentz boost,
but probably also by a transverse boost. It is in fact reasonable to
assume a distribution of transverse velocities of DCC such that the mean is
semirelativistic, say somewhere between 0.4 and 0.8. If this is the case,
and the internal relative velocities of the pions within the DCC cluster
are smaller or at least no larger, then it will often be the case that
the DCC in the laboratory frame will look like a coreless minijet. In
fact DCC searches within minijets, using the techniques sketched in the
next section, might be very fruitful.

In any case, these pieces of transversely boosted DCC, in reference frames where the
longitudinal velocity is zero or small, will typically occupy of order a
steradian of solid angle, indicating again that perhaps the natural
correlation length for DCC is of order 1--2.  However, there is an
unsolved theoretical issue here. Suppose one has a piece of DCC centered
at rapidity of  + 1, and another centered at $-$ 1 with a different chiral
order parameter.  They will have some
small overlap at rapidity 0. How do the two pieces interact? Will there
be a tendency to create a common alignment, or will they form independent
domains? I believe this to be an important fundamental question; it has
not yet received serious attention by theorists. The classical linear
sigma model should be sufficient as the fundamental theoretical tool; the
problem is the introduction of a realistic collision-geometry scenario
which is still computationally tractable.

There is even a remote possibility that the nonlinearities of the linear
sigma model are strong enough to promote long-range correlations in
rapidity, \ie\ to create a semiclassical structure which produces the
phenomenology of a soft Pomeron. For example it looks not at all out of
the question that a ladder built from pions on the sides and sigmas on
the rungs, with couplings dictated by the linear sigma model, could
produce a Regge intercept of unity or larger for forward hadron-hadron
scattering. But I do not know how such a Reggeon ladder could be related
to the classical DCC scenarios we have discussed.

\subsection{Charged DCC?}

Peter Lipa asked this question and, together with Brigitte Buschbeck, we
have begun to look into this issue\cite{refaa}. The idea is even more speculative
than ordinary DCC.  But if it makes any sense at all, it has the advantage
that one can make the search using data sets containing charged-particle
information only. This is not the case for ordinary DCC, where, as
discussed in the next section, the relationship between charged and
neutral pion production is what is examined. For charged DCC it is the
relationship between positive and negative pion production which is the
object of study.

Charged pion fields are to the real, Cartesian pion degrees of freedom
$\phi_1$ and $\phi_2$ as circularly polarized light is to linearly polarized
light. And certainly there are classical sources of circularly polarized light,
either associated with vorticity of the source, or with a 90$^\circ$ phase
difference of the sources of the two Cartesian components (or both).  So
from this point of view it seems not totally out of the question to
imagine a similar possibility for the pions. 

We are presently modeling a space-time scenario where DCC in the $\phi_1$ direction
is produced at positive eta, and DCC in the $\phi_2$ direction is produced
at negative eta, with a ``domain wall" in between.  We find, in the
spectrum of produced DCC pions,  a dipole
layer in the lego plot, with an average positive charge per pion on one
side of the domain wall, compensated by negative charge on the other. It
is to be emphasized that this is an average charge {\em per pion} at the
quantum level. The width of the dipole layer is 1--2 units of rapidity,
as might be anticipated on general grounds. We need to do a little more
work before reporting on its strength and momentum dependence.

\section{DCC Phenomenology}

\subsection{The inverse square root distribution.}

A basic signature of DCC production is the presence of very large
event-by-event fluctuations in the fraction of produced pions which are
neutral\cite{refaaa,refo,refl,refm}. Generic production models will give 
a distribution in the neutral fraction, defined as
\begin{equation}
f = \frac{N_{\pi^0}}{N_{\pi^0}+N_{\pi^+}+N_{\pi^-}}
\label{eq: V}
\end{equation}
which is binomial and peaked at $f = 1/3$. 
There will be in the classical large $N$ limit a very small probability
that, for example, all the pions are neutral. But if the pions are DCC
decay products this probability is not at all so small. 

The simplest estimate for the distribution of neutral fraction in DCC
production assumes that the chiral orientation in (Cartesian) isospin
space is random. Then a totally elementary calculation gives the result
that the distribution of neutral fraction is inverse-square-root.
With $ f = \cos^2\theta$ it follows that
\begin{equation}
 dn \approx d(\cos\theta) = \frac{1}{2\cos\theta}\
d(\cos^2\theta) = \frac{df}{2\sqrt f} \ .
\label{eq:W }
\end{equation}
It is probable however that this component is to some extent immersed in
generic background. Cuts in $p_t$, for example, may be useful in enhancing
the signal. 

It is also the case that it appears to be better to use an indirect
technique to test for the presence of the inverse-square-root component\cite{refbb}.
This utilizes the machinery of multiparticle production dynamics and
multiplicity distributions, to which we now turn. 

\subsection{Multiplicity distributions and generating functions}

In the MiniMax experiment, the phase-space coverage is small, about 1.0
lego-area units. For that analysis it is reasonable to assume that within
the acceptance the chiral order parameter takes a fixed value.

The raw information relevant to DCC physics is the multiplicity
distribution of produced pions. Ideally they should be momentum analyzed,
with the data binned in intervals of $p_t$. However this so far has been not
done experimentally, and we simplify by ignoring the momentum degree of
freedom. Let the probability of producing $N$ pions be $P(N)$, and introduce
the generating function
\begin{equation}
G(z) = \sum_N z^NP(N)
\label{eq: X}
\end{equation}
which contains all information about the multiplicity distribution. For a
Poisson distribution, the generating function is an exponential
\begin{equation}
G(z) = e^{\mu(z-1)} \qquad P(N) = \frac{\mu^Ne^{-\mu}}{N!} \ .
\label{eq: Y}
\end{equation}
In the general case it is a superposition
of Poisson distributions with positive semidefinite weight
function $\rho$ 
\begin{equation}
G(z) = \int^\infty_0 d\mu\, \rho(\mu)\, e^{\mu(z-1)} \ .
\label{eq: Z}
\end{equation}

For the two species of charged and neutral pions, the generalization is a
generating function of two variables, again a superposition of Poisson
distributions. The definition of generic pion production is that the only
correlation is produced by the aforementioned $\rho$ for the total pion
multiplicity distribution, so that
\begin{eqnarray}
G(z_{ch},z_0) &=& \sum P(N_{ch},N_0) z^{N_{ch}}_{ch}z^{N_0}_0 
\nonumber\\
&=& \int^\infty_0 d\mu\, \rho(\mu)\, e^{\mu\left[f(z_0-1)+(1-f)(z_{ch}-1)\right]}
\label{eq: aa}
\end{eqnarray}
with the neutral fraction $f$ approximately 1/3.

By expanding things out, one sees that the partition into charged and
neutral pions is governed by a binomial distribution; indeed this could
have been the common-sense starting point. It is also the case that
existing Monte-Carlo codes have the property that the distribution of the
neutral fraction is approximately binomial. 

For pure DCC, all one needs do is introduce the inverse square-root
distribution in $f$ as another weight factor:
\begin{equation}
\VEV{\ } \equiv \int^\infty_0 d\mu\, \rho(\mu) \Rightarrow \int^\infty_0 
d\mu\, \rho(\mu)\int^1_0 \frac{df}{2\sqrt f} \ .
\label{eq:bb }
\end{equation}
One sees that the basic difference between generic production and DCC
production is that in the former case the generating function depends only
upon one variable, while for DCC it depends nontrivially upon two. A very
good way of testing for the distinction is via factorial moments\cite{refcc}. These
are just the derivatives of the generating function with respect to the
$z$'s at $z = 1$. In the case of only one variable, one has
\begin{equation}
f_1 = \VEV n = \left.\frac{\partial G}{\partial z}\right|_{z=1} \qquad
f_2 = \VEV{n(n-1)} = \left.\frac{\partial^2G}{\partial z^2}\right|_{z=1} \ .
\label{eq:cc }
\end{equation}
The normalized factorial moments are
\begin{equation}
F_n = \frac{1}{\VEV n^n}\
\left.\frac{\partial^2G}{\partial z^n}\right|_{z=1} = \frac{f_n}{(f_1)^n} \ .
\label{eq:dd }
\end{equation}
In the case of interest, there is a two-dimensional array of normalized
factorial moments which captures the information contained in the joint
multiplicity distribution: 
\begin{equation}
F_{ij} = \frac{1}{\VEV{n_{ch}}^i\VEV{n_0}^j} \
\frac{\partial^{i+j}G}{\partial z^i_{ch}\partial z_0^j} \ .
\label{eq:ee }
\end{equation}
However for generic production there are many relations between these,
because the generating function depends upon only one variable: 
\begin{equation}
F_{ij} = F_{(i+j),0} \ .
\label{eq:ff }
\end{equation}
Therefore many ratios of the $F_{ij}$ are expected to be unity
\begin{equation}
r_{ij} = \frac{F_{ij}}{F_{(i+j),0}} = 1 \qquad
\mbox{(Generic Production)}
\label{eq:gg }
\end{equation}
while for pure DCC they can also be explicitly computed, and are far from
unity:
\begin{equation}
r_{ij} = \frac{F_{ij}}{F_{(i+j),0}} =
\frac{\VEV{(1-f)^if^j}\VEV{(1-f)^j}}
{\VEV{(1-f)^{i+j}}\VEV{f^j}} =
\frac{i!(2j-1)!!}{(i+j)!} \ .
\label{eq: hh}
\end{equation}

There are also experimental reasons why these ratios are useful.  First of
all, one can go from $\pi^0$ production to photon production via convolution,
and the factorial moment method remains robust: one simply replaces the
neutral-pion fugacity $z_0$ by the generating function for the gammas, a
second-order polynomial in the photon fugacity. In addition, if
efficiencies are not 100 percent, but are uncorrelated with total
multiplicity or other global parameters, they can be incorporated in terms
of modified fugacities. The properties of the resultant factorial moments,
which are the direct observables experimentally, essentially do not
change. That is, the bivariate factorial moment ratios directly extracted
from data on production of charged hadrons and of gammas, as in Eqs.(33)
and (34), still will be unity for generic production and far from unity in
the case of DCC production. 

\subsection{Doing better: correlation functions and generating functionals}

Once the experimental acceptance becomes large, or if one investigates,
\eg\ the $p_t$ dependence of the presumed DCC fraction (large at low $p_t$,
small at high $p_t$?), then the correlation structure of the DCC order
parameter, and even the generic particle distributions themselves, becomes
of paramount importance. Clearly the formalism of generating functions
should be preserved as closely as possible, and this is in principle a
straightforward matter of replacing generating functions with generating
functionals. 

In the case of DCC production, the bridge from the classical calculations
to phenomenology is fairly clear. For a given classical solution of, say,
the linear sigma model there will be a source function
\begin{equation}
(\Box+\mu^2)\, \phi(x) \equiv J(x) \ .
\label{eq: ii}
\end{equation}
As we discussed, the multiplicity distribution is built from the squared
Fourier transform of the source function, put on mass shell,
\begin{equation}
2E \, \frac{dN}{d^3p} = \frac{1}{(2\pi)^3} \left|\tilde J(p)\right|^2 \ .
\label{eq:jj }
\end{equation}
The fluctuations for the classical case are Poissonian, and so the
generating functional is again just an exponential in all the continuous
number of fugacity variables, now parametrized by the particle momenta. 
Finally one should average over the choice of source function, which at
the least is parametrized by choices of initial conditions for the
classical field configurations (including chiral orientation). This leads
to a DCC generating functional of the form
\begin{equation}
G_{DCC}=\int \D J_1\ \D J_2\ \D J_3\ \P(J)\exp\sum_i
\frac{d^3p}{2E(2\pi)^3}\
\left|\tilde J_i(p)\right|^2(z_i(p)-1) \ . 
\label{eq:kk }
\end{equation}
A generalized DCC distribution of the inverse square root type follows if
\begin{equation}
\P(J) \Rightarrow \P(|\vec J|^2) \ .
\label{eq:ll }
\end{equation}
Thus far this generalization is reasonably straightforward, and leaves the
unknown issues mostly at the level of the choice of boundary conditions
and structure of the classical field equations considered in the previous
section. However it is not reasonable to assume that the totality of
particle production originates as DCC. And there is no consensus on what
generating functional to use to describe generic particle production. Many
correlation phenomena exist, some from minijets and perturbative QCD,
undoubtedly some associated with the fluctuations in the number of
``wounded" constituent quarks per collision, still others associated with
impact parameter dependence, and more associated with resonance
production. In addition, one can consider various ways of combining the
generic particle production with DCC production. Three extreme cases are
as follows: 

\bigskip
\noindent
1)
\begin{equation}
G = G_{\rm generic} + G_{DCC} \ .
\label{eq:mm }
\end{equation}
This means that in a given event either generic particles are
produced or DCC, but not both.

\bigskip
\noindent
2)
\begin{equation}
G = G_{\rm generic} \cdot G_{DCC} \ .
\label{eq:nn }
\end{equation}

In this case the amount of DCC produced in a given event is not
correlated at all with the amount of generic particles. 

\bigskip
\noindent
3)
\begin{equation}
z = z_{\rm generic} + z_{DCC} \ .
\label{eq:oo }
\end{equation}
In this case the amount of DCC produced is, up to binomial-distribution fluctuations, 
in proportion to the amount of generic particle production. 

The phenomenological consequences of this distinction are very strong; a
small DCC admixture of the first type is much easier to isolate
experimentally than a small admixture of the third type. The case of
independent production is intermediate.

\subsection{DCC vs. Bose-Einstein, etc.}

There is an interesting question of whether observation of the inverse
square-root distribution implies observation of Baked-Alaska DCC.  This is
not at all clear. The inverse square-root distribution was discovered in a
different context, namely the production of a cluster of pions in a
maximally symmetric state. The inverse square-root behavior is also a
consequence of the well-known Andreev, Plumer, Weiner (APW)  
description\cite{refee}
of Bose-Einstein correlations. In addition, it has been shown that in the
tree-level expansion of the chiral effective theory, the inverse-square
root distribution emerges\cite{refff}.
But none of these descriptions follows the same line as the DCC
description above. In particular, the part of the APW scenario
which leads to DCC-like behavior assumes a random Gaussian
distribution of source functions, except with respect to the isospin
degrees of
freedom. In the DCC description, one computes the sources via the
sigma-model equations of motion and (most probably) a random set of
initial conditions. This would for sure make the distribution of
DCC sources
non-random functionals. And the tree-level chiral expansion does not seem
to have the possibility of incorporating the Baked-Alaska physics. 

It is probably the case that one should view DCC as a specific mechanism
of Bose-Einstein enhancement. There seems to me nothing wrong with this
point of view, because there really is an ongoing problem of describing
Bose-Einstein effects from the most general point of view. In any case,
for me this is not a very important issue until there is concrete
experimental evidence
for DCC-like behavior. First of all something needs to be seen. Thereafter
there will be plenty of opportunity for figuring out what it means via
further interaction between follow-on observations and the theory.

\subsection{Charged DCC again}

The phenomenological description of the charged DCC production discussed
in Section 2.8 is similar to that of the conventional DCC. It is being
worked out by Peter Lipa, Brigitte Buschbeck\cite{refaa}, and myself.

The major change is that the source function in momentum space is taken to
be complex
\begin{equation}
J_i = u_i + iv_i 
\label{eq:pp }
\end{equation}
and the real and imaginary parts $u$ and $v$ are each taken as random
variables. In the usual case there is assumed to be complete correlation
between $u$ and $v$. One finds

\bigskip
\noindent
Before: 
\begin{equation}
G = \int \D^3J\, f(|J|^2)\, e^F \Rightarrow
\int \D J|J|^2 f(|J|^2) \int^1_0\frac{df}{2\sqrt f}\
e^F\label{eq:qq }
\end{equation}
After: 
\begin{equation}
G = \int \D^3u\ \D^3v\, f(|J|^2)\, e^F\Rightarrow
\int \D J|J|^2f(|J|^2)
\int^1_0df(1-f)\int^1_{-1}dq\, e^F\label{eq: rr}
\end{equation}
with
\begin{equation}
q = \frac{\VEV{(N_+-N_-)}}{\VEV{(N_++N_-)}} \ .
\label{eq:ss }
\end{equation}
For the ``action'' of the generating function one has 
\begin{equation}
F = \mu\left[(z_+-1)\, \left(\frac{1+q}{2}\right)\, (1-f) + (z_--1)\,
\left(\frac{1-q}{2}\right)\, (1-f) + (z_0-1)\, f\right] \ .
\label{eq: tt}
\end{equation}
Again the MiniMax-like ratios of factorial moments can be constructed, and
compared with charged particle data. It appears from a cursory first look
that when averaged over all $p_t$ the data will be of opposite sign to the
charged-DCC expectation. Charge tends to be locally neutralized in the
lego plot, more than it would from a random throwing of charge into phase
space. This to be expected in both the string and parton cascade pictures;
there is very little charge separation and flow in the space-time
evolution. 

However when the $p_t$'s are low, the experimental situation changes. It is
here that the charged-DCC hypothesis is not ruled out, and the detailed
phenomenology of this region of phase-space may turn out to be
interesting. 

The assumption of complete randomness of $u$ and $v$ is an extreme one. There
are evidently interpolations between the extremes of no correlation and
complete correlation which can be easily constructed.  Study of the
structure of the classical models as outlined in Section 2.8 will help in
choosing a reasonable starting hypothesis.

\section*{Acknowledgments}

Thanks go to Jan Czyzewski, Jacek Wosiek, and all the organizers of the
Zakopane School for a most productive and pleasant meeting. It is also a
great pleasure to dedicate these notes to my good friend and colleague
Wieslaw Czyz on the occasion of his seventieth birthday.

\end{document}